# EXTRA DIMENSIONS, BRANE WORLDS, AND THE VANISHING OF AXION CONTRIBUTIONS TO INFLATION?

## A. W. Beckwith

## ABSTRACT

We examine from first principles the implications of the $5^{th}$ Randall-Sundrum Brane world dimension in terms of setting initial conditions for chaotic inflationary physics. Our model pre supposes that the inflationary potential pioneered by Guth is equivalent in magnitude in its initial inflationary state to the effective potential presented in the Randall - Sundrum model We also consider an axion contribution to chaotic inflation (which may have a temperature dependence) which partly fades out up to the point of chaotic inflation being matched to a Randall – Sundrum effective potential. If we reject an explicit axion mass drop off to infinitesimal values at high temperatures, we may use the Bogomolnyi inequality to re scale and re set initial conditions for the chaotic inflationary potential. One of potential systems embedded in the Randall - Sundrum brane world is a model with a phase transition bridge from a tilted washboard potential to the chaotic inflationary model pioneered by Guth which is congruent with the slow roll criteria. If, as written up earlier, the axion wall contribution is due to di-quarks, this is equivalent to tying in baryogenesis to the formation of chaotic inflation initial conditions, with the Randall - Sundrum brane world effective potential delineating the end of the dominant role of di-quarks, and the beginning of inflation.

Correspondence: A. W. Beckwith:   projectbeckwith2@yahoo.com

PACS numbers: 95.10.-a, 98.80.-k, 98.80.Cq

**INTRODUCTION**

This investigation is attempting to show that the fifth dimension postulated by Randall-Sundrum theory helps give us an action integral which leads to a minimum physical potential we can use to good effect in determining initial conditions for the onset of inflation. The 5$^{th}$ dimension of the Randall-Sundrum brane world is of the genre, for $-\pi \leq \theta \leq \pi$

$$x_5 \equiv R \cdot \theta \tag{1}$$

This lead to an additional embedding structure for typical GR fields, assuming as one may write up a scalar potential 'field 'with $\phi_0(x)$ real valued, and the rest of it complex valued as[1]:

$$\phi(x^\mu, \theta) = \frac{1}{\sqrt{2 \cdot \pi \cdot R}} \cdot \left\{ \phi_0(x) + \sum_{n=1}^{\infty} [\phi_n(x) \cdot \exp(i \cdot n \cdot \theta) + C.C.] \right\} \tag{2}$$

This scalar field makes its way to an action integral structure which will be discussed later on, which Sundrum used to forming an effective potential. Our claim in this analysis can also be used as a way of either embedding a Bogomolyni inequality, perhaps up to five dimensions[2], or a straight forward reduction in axion mass due to a rise in temperature[3] helped reduced effective potential in this structure, with the magnitude of the Sundrum potential forming an initial condition for the second potential of the following phase transition .Note that we are referring to a different form of the scalar potential, which we will call $\tilde{\phi}$, which has the following dynamic[4].

$$\begin{array}{ll} \tilde{V}_1 & \rightarrow \tilde{V}_2 \\ \tilde{\phi}(increase) \leq 2 \cdot \pi & \rightarrow \tilde{\phi}(decrease) \leq 2 \cdot \pi \\ t \leq t_P & \rightarrow t \geq t_P + \delta \cdot t \end{array} \tag{3}$$

The potentials $\tilde{V}_1$, and $\tilde{V}_2$ were described in terms of **S-S'** di quark pairs nucleating and then contributing to a chaotic inflationary scalar potential system. Here, $m^4 \approx (1/100) \cdot M_P^4$

$$\tilde{V}_1(\phi) = \frac{M_P^4}{2} \cdot (1 - \cos(\tilde{\phi})) + \frac{m^4}{2} \cdot (\tilde{\phi} - \phi^*)^2 \tag{3a}$$

$$\tilde{V}_2(\phi) \propto \frac{1}{2} \cdot (\tilde{\phi} - \phi_C)^2 \tag{3b}$$

We should keep in mind that $\phi_C$ in Eqn 3a is an equilibrium value of a true vacuum minimum of Eqn. 3a after tunneling. In the potential system given as Eqn, (3a) we see a steadily rising scalar field value which is consistent with the physics of Figure 1 . In the potential system given by Eqn. (3b) we see a reduction of the 'height of a scalar field which is consistent with the chaotic inflationary potential overshoot phenomena We should note that $\phi^*$ in Eq (3a )is a measure of the onset of quantum fluctuations. **Appendix I** is a discussion of Axion potentials which we claim is part of the contribution of the potential given in Eqn. (3a) Note that the tilt to the potential given in Eqn. (3a) is due to a quantum fluctuation. As explained by Guth for quadratic potentials[5],

$$\phi^* \equiv \left(\frac{3}{16 \cdot \pi}\right)^{\frac{1}{4}} \cdot \frac{M_P^{3/2}}{m^{\frac{1}{2}}} \cdot M_P \rightarrow \left(\frac{3}{16 \cdot \pi}\right)^{\frac{1}{4}} \cdot \frac{1}{m^{\frac{1}{2}}} \tag{3c}$$

This in the context of the fluctuations having an upper bound of

$$\tilde{\tilde{\phi}} > \sqrt{\frac{60}{2\cdot\pi}} M_P \approx 3.1 M_P \equiv 3.1 \tag{3d}$$

Here, $\tilde{\tilde{\phi}} > \phi_C$. Also, the fluctuations Guth had in mind were modeled via[6]

$$\tilde{\phi} \equiv \tilde{\tilde{\phi}} - \frac{m}{\sqrt{12\cdot\pi\cdot G}} \cdot t \tag{3e}$$

In the potential system given by Eqn. (3b) we see a reduction of the 'height' or magnitude of a scalar field which is consistent with the chaotic inflationary potential overshoot phenomena mentioned just above. This leads us to use the Randal-Sundrum effective potential[1], in tandem with tying in **baryogenesis**[7] to the formation of chaotic inflation initial conditions for Eqn. (3b), with the Randall-Sundrum brane world effective potential delineating the end of the dominant role of di quarks, due to baryogenesis, and the beginning of inflation. The role of the Bogomolnyi inequality is to introduce, from a topoplogical domain wall stand point a mechanism for the introduction of baryogenesis in early universe models, and the combination of that analysis, plus matching conditions with the Randal-Sundrum effective potential sets us up for chaotic inflation.

## HOW TO FORM THE RANDALL-SUNDRUM EFFECTIVE POTENTIAL

The consequences of the fifth dimension mentioned in Eqn. (1) above show up in a simple warped compactification involving two branes, i.e. a Planck world brane, and an IR brane. This construction with the physics of this 5 dimensional system allow for solving the hierarchy problem of particle physics, and in addition permits us to investigate the following five dimensional action integral.[1]

$$S_5 = \int d^4x \cdot \int_{-\pi}^{\pi} d\theta \cdot R \cdot \left\{ \frac{1}{2} \cdot (\partial_M \phi)^2 - \frac{m_5^2}{2} \cdot \phi^2 - K\cdot\phi\cdot[\delta(x_5) + \delta(x_5 - \pi\cdot R)] \right\} \tag{4}$$

This integral, will lead to the following equation to solve

$$-\partial_\mu \partial^\mu \phi + \frac{\partial_\theta^2}{R^2}\phi - m_5^2\phi = K\cdot\frac{\delta(\theta)}{R} + K\cdot\frac{\delta(\theta-\pi)}{R} \tag{5}$$

Here, what is called $m_5^2$ can be linked to Kalusa Klein "excitations"[1] via (for $n > 0$)

$$m_n^2 \equiv \frac{n^2}{R^2} + m_5^2 \tag{6}$$

This uses[8] (assuming $l$ is the curvature radius of AdS$_5$)

$$m_5^3 \equiv \frac{M_P^2}{l} \tag{6a}$$

This is for a compactification scale, for $m_5 \ll \frac{1}{R}$, and after an ansatz of the following is used:

$$\phi \equiv A \cdot [\exp(m_5 \cdot R \cdot |\theta|) + \exp(m_5 \cdot R \cdot (\pi - |\theta|))] \tag{7}$$

We then obtain after a non trivial vacuum averaging

$$\langle \phi(x,\theta) \rangle = \Phi(\theta) \tag{8}$$

$$S_5 = -\int d^4x \cdot V_{eff}(R_{phys}(x)) \tag{9}$$

This is leading to an initial formulation of

$$V_{eff}(R_{phys}(x)) = \frac{K^2}{2 \cdot m_5} \cdot \frac{1 + \exp(m_5 \cdot \pi \cdot R_{phys}(x))}{1 - \exp(m_5 \cdot \pi \cdot R_{phys}(x))} \tag{10}$$

Now, if one is looking at an addition of a 2$^{nd}$ scalar term of opposite sign, but of equal magnitude[1]

$$S_5 = -\int d^4x \cdot V_{eff}(R_{phys}(x)) \rightarrow -\int d^4x \cdot \tilde{V}_{eff}(R_{phys}(x)) \tag{11}$$

This is for when we set up an effective Randall – Sundrum potential looking like[1]

$$\tilde{V}_{eff}(R_{phys}(x)) = \frac{K^2}{2 \cdot m_5} \cdot \frac{1 + \exp(m_5 \cdot \pi \cdot R_{phys}(x))}{1 - \exp(m_5 \cdot \pi \cdot R_{phys}(x))} + \frac{\tilde{K}^2}{2 \cdot \tilde{m}_5} \cdot \frac{1 - \exp(\tilde{m}_5 \cdot \pi \cdot R_{phys}(x))}{1 + \exp(\tilde{m}_5 \cdot \pi \cdot R_{phys}(x))} \tag{12}$$

This above system has a meta stable vacuum for a given special value of $R_{phys}(x)$ We will from now on use this as a 'minimum' to compare a similar action integral for the potential system given by Eqn. (3) above. Note that this is done, while assuming that

## HOW TO COMPARE THE RANDALL-SUNDRUM EFFECTIVE POTENTIAL MINIMUM WITH AN EFFECTIVE POTENTIAL MINIMUM INVOLVING THE POTENTIAL OF EQN. (3) ABOVE.

We are forced to consider two possible routes to the collapse of a complex potential system to the chaotic inflationary model promoted by Guth[5].

The first such model involves a simple reduction of the axion wall potential[9] as given by, especially when N = 1

$$V(a) = m_a^2 \cdot (f_{PQ}/N)^2 \cdot (1 - \cos[a/(f_{PQ}/N)]) \tag{13}$$

The simplest way to deal with Eqn.(13) is to set $m_a^2(T) \xrightarrow[T \to \infty]{} \varepsilon^+$, when Kolb[9] writes

$$m_{axion}(T) \cong .1 \cdot m_{axion}(T=0) \cdot (\Lambda_{QCD}/T)^{3.7} \tag{14}$$

i.e. to declare that the axion 'mass' vanishes, and to let this drop off in value give a simple truncated version of chaotic inflationary potentials along the lines given by a transition from Eqn (3a) to Eqn. (3b) We should note that $\Lambda_{QCD}$ is the enormous value of the cosmological constant which is $10^{120}$ larger than what it is observed to be today[10,11,12], and for now we are side steeping the question of if or not the negative valued Randall-Sundrum cosmological constant[8]

$$\Lambda_5 = -\frac{6}{l^2} \tag{15}$$

has a bearing on this situation. Not to mention the problems inherent in several proposed fixes to the cosmological constant problem[13].

Now if we want an equivalent explanation, which may involve baryogenesis, we need to look at the component behavior of each of the terms in Eqn. (13) without assuming $m_a^2(T) \xrightarrow[T \to \infty]{} \varepsilon^+$. Then, we need to re define several of the variables presented above. Now, in the typical theory presented by

$$\frac{M_P^4}{2} \cdot (1 - \cos(\tilde{\phi})) \propto m_a^2 \cdot (f_{PQ}/N)^2 \cdot (1 - \cos[a/(f_{PQ}/N)]) \tag{16}$$

We then have to present a varying in magnitude value for the 'scalar' $\tilde{\phi}$ involving ultimately the Bogolmolnyi inequality. I have done several of these for condensed matter current problems, but for our cosmology situation, we first have to work with

$$[a/(f_{PQ}/N)] \approx \tilde{\phi} \tag{17}$$

There has been credible work with instantons in higher dimensions, starting with Hawkings 1999 article[14] This, however, addresses a way of linking an instanton structure with baryogenesis, dark energy, and issues of how Randall-Sundrum brane structrure can be used to formation of initial conditions of inflationary cosmology.

Clarifying what can be done with an instanton style quantum nucleation in multiple dimensions[15] may help us with more acceptable models[16,7] as to estimating, roughly, a quantum value for the cosmological constant, as an improvement in recent calculations.

$$\Psi \propto \exp(-\int d^3 x_{space} d\tau_{Euclidian} L_E) \equiv \exp\left(-\int d^4 x \cdot L_E\right) \tag{18}$$

$$L_E \geq |Q| + \frac{1}{2} \cdot (\tilde{\phi} - \phi_0)^2 \{ \} \xrightarrow[Q \to 0]{} \frac{1}{2} \cdot (\tilde{\phi} - \phi_0)^2 \cdot \{ \} \tag{18a}$$

Where

$$\{ \} = 2 \cdot \Delta \cdot E_{gap} \tag{18b}$$

This leads, if done correctly to the quadratic sort of potential contribution as given by[16], At the same time it raises the question of if or not when there is a change from the 1st to the 2nd potential system, we have a consistent model. Let us now view a toy problem involving use of a S-S' pair which we may write as[17]

$$\tilde{\phi} \approx \pi \cdot [\tanh b(x - x_a) + \tanh b(x_b - x)] \tag{19}$$

This is for a di quark pair along the lines given when looking at the first potential system, which is a take off upon Zhitinisky's color super conductor model[18]

**COMPARISON OF INITIAL CONDITIONS FOR A NUCLEATING UNIVERSE.**

Now for the question the paper is raising., Can we realistically state the following for initial conditions of a nucleating universe ? If so, then what are the consequences?

$$S_5 = -\int d^4 x \cdot \tilde{V}_{eff}\left(R_{phys}(x)\right) \propto \left(-\int d^3 x_{space} d\tau_{Euclidian} L_E\right) \equiv \left(-\int d^4 x \cdot L_E\right) \tag{20}$$

The right hand side of Eqn (20) can be stated as having

$$L_E \geq \frac{1}{2} \cdot \left(\tilde{\phi} - \phi_0\right)^2 \cdot \{\ \}. \tag{21}$$

We can insist that this $\Delta E_{gap}$ between a false and a true vacuum minimum[19], that

$$\{\ \} \equiv 2 \cdot \Delta E_{gap} \tag{22}$$

So, this leads to the following question. Does a reduction of axion wall mass for the first potential system given in Eqn.(3a) being transformed to Eqn.(3b) above give us consistent physics, due to temperature dependence in axion 'mass', or should we instead look at what can be done with S-S' instanton physics and the Bogolmyi inequality[20], in order to perhaps take into account Baryogenesis ? Also, can this shed light upon the Wheeler De Witts equations[21] modification by Ashtekar [22] in early universe quantum bounce conditions ?

Finally, does this process of baryogenesis,if it occurs lend then to the regime where there is a bridge between classical applications of the Wheeler De Witt equation to the quantum bounce condition raised by Ashtekar [22]?

### TIE IN WITH THE WHEELER DE-WITT EQUATION.

Abbay Ashtekar's quantum bounce[22] gives a discretized version of the Wheeler De Witt equation. Let us first review classical De Witt theory which incidently ties in with inflationary n= 2 scalar potential field cosmology.This will be useful in analyzing consequences of the wave functional so formed in Eqn. (18) and suggest quantum bounce analogies we will comment upon later.

In the common versions of Wheeler De Witt theory a potential system using a scale radius $R(t)$, with $R_0$ as a classical turning point value[21]

$$U(R) = \left(\frac{3 \cdot \pi \cdot c^3 \cdot R_0}{2 \cdot G}\right)^2 \cdot \left[\left(\frac{R}{R_0}\right)^2 - \left(\frac{R}{R_0}\right)^4\right] \tag{23}$$

Here we have that

$$R_0 \sim c \cdot t_0 \equiv l_P \equiv c \cdot \sqrt{\frac{3}{\Lambda}} \sim 7.44 \times 10^{-36}\, meters \tag{23a}$$

As well as

$$\sqrt{\frac{3}{\Lambda}} \equiv t_P \sim 2.48 \times 10^{-44}\, \sec \tag{24}$$

Now, Alfredo B. Henriques[16] presents a way in which one can obtain a Wheeler De Witt equation based upon

$$\tilde{\hat{H}} \cdot \Psi(\phi) = \left[ \frac{1}{2} \cdot \left( A_\mu \cdot p_\phi^2 + B_\mu \cdot m^2 \cdot \phi^2 \right) \cdot \Psi(\phi) \right] \tag{25}$$

Using a momentum operator as give by

$$\hat{p}_t = -i \cdot \hbar \cdot \frac{\partial}{\partial \cdot \phi} \tag{26}$$

This is assuming a real scalar field $\phi$ as well as a 'scalar mass' $m$ 'based upon a derivation originally given by Thieumann[23]. The above equation as given by Theumann, and secondarily by Henriques[16] lead directly to considering the real scalar field $\phi$ as leading to a prototype wave functional for the $\phi^2$ potential term as given by

$$\psi_\mu(\phi) \equiv \psi_\mu \cdot \exp(\alpha_\mu \cdot \phi^2) \tag{27}$$

As well as an energy term

$$E_\mu = \sqrt{A_\mu \cdot B_\mu} \cdot m \cdot \hbar \tag{27a}$$

$$\alpha_\mu = \sqrt{B_\mu / A_\mu} \cdot m \cdot \hbar \tag{27b}$$

This is for a 'cosmic' Schrodinger equation as given by

$$\tilde{\hat{H}} \cdot \psi_\mu(\phi) = E_\mu(\phi) \tag{27c}$$

This has

$$A_\mu = \frac{4 \cdot m_{pl}}{9 \cdot l_{pl}^9} \cdot \left( V_{\mu+\mu_0}^{1/2} - V_{\mu-\mu_0}^{1/2} \right)^6 \tag{27d}$$

And

$$B_\mu = \frac{m_{pl}}{l_{pl}^3} \cdot (V_\mu) \tag{27e}$$

Here $V_\mu$ is the eignvalue of a so called volume operator[16] and the interested readers are urged to consult with the cited paper to go into the details of this, while at the time noting $m_{pl}$ is for Planck mass, and $l_{pl}$ is for Planck length, and keep in mid that the main point made above, is that a potential operator based upon a quadratic term leads to a Gaussian wavefunctional with an exponential similarly dependent upon a quadratic $\phi^2$ exponent. We do approximate solitons via the evolution of Eqn. (27) and Eqn. (27c) above, and so how we reconcile higher order potential terms in this approximation of wave functionals is extremely important.

Now Ashtekar in his longer arXIV article[24] make reference to a revision of this momentum operation along the lines of basis vectors $|\mu\rangle$ by

$$\hat{p}_t |\mu\rangle = \frac{8 \cdot \pi \cdot \gamma \cdot l_{PL}^2}{6} \cdot \mu |\mu\rangle \tag{28}$$

With the advent of this re definition of momentum we are seeing what Ashtekar works with as a sympletic structure with a revision of the differential equation assumed in Wheeler – De Witt theory to a form characterized by[26]

$$\frac{\partial^2}{\partial \phi^2} \cdot \Psi \equiv - \Theta \cdot \Psi \tag{28a}$$

$\Theta$ in this situation is such that

$$\Theta \neq \Theta(\phi) \tag{28b}$$

Also, and more importantly this $\Theta$ is a difference operator, allowing for a treatment of the scalar field as an 'emergent time', or 'internal time' so that one can set up a wave functional built about a Gaussian wavefunctional defined via

$$\max \tilde{\Psi}(k) = \tilde{\Psi}(k)\big|_{k \equiv k^*} \tag{28c}$$

This is for a crucial 'momentum' value

$$p_\phi^* = - \left(\sqrt{16 \cdot \pi \cdot G \cdot \hbar^2 / 3}\right) \cdot k^* \tag{28d}$$

And

$$\phi^* = -\sqrt{3/16 \cdot \pi G} \cdot \ln|\mu^*| + \phi_0 \tag{29}$$

Which leads to, for an initial point in 'trajectory space' given by the following relation $(\mu^*, \phi_0) =$ (initial degrees of freedom [dimensionless number] ~'eignvalue of 'momentum', initial 'emergent time ' )
So that if we consider eignfunctions of the De Witt (difference) operator, as contributing toward

$$e_k^s(\mu) = \left(1/\sqrt{2}\right) \cdot [e_k(\mu) + e_k(-\mu)] \tag{30a}$$

With each $e_k(\mu)$ an eignfunction of Eqn. (12a) above, with eignvalues of Eqn. (12a) above given by $\omega(k)$, we have a potentially numerically treatable early universe wave functional data set which can be written as

$$\Psi(\mu, \phi) = \int_{-\infty}^{\infty} dk \cdot \tilde{\Psi}(k) \cdot e_k^s(\mu) \cdot \exp[i\omega(k) \cdot \phi] \tag{30b}$$

This equation above has a 'symmetry' as seen in Figure 1 of Ashtekar's PRL article [6] about $\phi$, reflecting upon a quantum bounce for a pre ceding universe prior to the 'big bang' contracting to the singularity and a 'rebirth ' as seen by a different 'branch of Eqn. (30b) emerging for a 'growing' set of values of $\phi$..

## CONCLUSION.

We are presenting a question which may be of relevance to JDEM research. Namely if Ashtekar is correct in his quantum geometry[24], and the break down of early universe conditions not permitting the typical application of the Wheeler De Witt equation, then what do we have to verify it experimentally? The axion wall dependence so indicated above may provide an answer to that, and may be experimentally measurable via Kadotas pixel reconstructive scheme.[25]

Furthermore, we also argue that the semi classical analysis of the initial potential system as given by Eqn (**3**) above and its subsequent collapse is de facto evidence for a phase transition to conditions allowing for dark energy to be created at the beginning of inflationary cosmology..[26,27]. This builds upon an earlier paper done by Kolb in minimum conditions for reconstructing scalar potentials[28,29,30]. It also will necessitate reviewing other recent derivation bound to the cosmological constant in cosmology model in a more sophisticated manner than has been presently done[31] In doing so, it may be appropriate to try to reconcile A. Ashtekar's approach involving a discretization of the Wheeler De Witt equation with the bounce calculations in general cosmology pioneered by Hackworth and Weinberg[32]… Needless to say, the work so presented above leaves open the question if or not baryogenesis, is involved in involving a collapse of the first term of Eqn. (3a) along the lines of the Bogomolnyi inequality, or else we have to skip this and to adhere to the topological defect models pioneered by Trodden,et al[33].

## APPENDIX I.

### FORMING AN AXION POTENTIAL TERM AS PART OF THE CONTRIBUTION TO EQUATION 2A

Kolb's book[7] has a discussion of an Axion potential given in his Eqn. (10.27)

$$V(a) = m_a^2 \cdot (f_{PQ}/N)^2 \cdot (1 - \cos[a/(f_{PQ}/N)]) \tag{1}$$

Here, he has the mass of the Axion potential as given by $m_a$ as well as a discussion of symmetry breaking which occurs with a temperature $T \approx f_{PQ}$. Furthermore, he states that the Axion goes to a massless regime for high temperatures, and becomes massive as the temperature drops. Due to the fact that Axions were cited by Zhitinisky in his QCD ball formation[18] this is worth considering, and I claim that this potential is part of Eqn. (6a) with the added term giving a tilt to this potential system, due to the role quantum fluctuations play in inflation. Here, N>1 leads to tipping of the wine bottle potential, and N degenerate CP-conserving minimal values. The interested reader is urged to consult section 10.3 of Kolb's Early universe book for additional details[9].